\pdfoutput=1
\documentclass[12pt]{iopart}

\usepackage{amsmath}
\usepackage{graphics}
\usepackage{bbm,color,calc}
\usepackage[hidelinks]{hyperref}
\usepackage[activate=normal]{pdfcprot}

\newcommand{\bfalpha}{\boldsymbol{\alpha}}
\newcommand{\bfgamma}{\boldsymbol{\gamma}}
\newcommand{\bfzero}{\mathbf{0}}
\newcommand{\bfe}{\mathbf{e}}
\newcommand{\bfh}{\mathbf{h}}
\newcommand{\bfM}{\mathbf{M}}\newcommand{\bfm}{\mathbf{m}}
\newcommand{\bfP}{\mathbf{P}}\newcommand{\bfp}{\mathbf{p}}
\newcommand{\bfr}{\mathbf{r}}
\newcommand{\bfs}{\mathbf{s}}
\newcommand{\bfu}{\mathbf{u}}
\newcommand{\bfV}{\mathbf{V}}

\newcommand{\calF}{\mathcal{F}}
\newcommand{\gulli}{UMR Gulliver 7083 CNRS, ESPCI ParisTech,
PSL~Research University, 10~rue~Vauquelin, 75005~Paris, France}

\begin{document}

\title{Coupling spin to velocity: collective motion of Hamiltonian polar particles}

\author{Sigbj{\o}rn L{\o}land Bore$^1$, Michael Schindler$^1$,
Khanh-Dang Nguyen Thu Lam$^1$, Eric Bertin$^{2,3}$, and Olivier Dauchot$^1$}
\address{$^1$ \gulli}
\address{$^2$ Universit\'e Grenoble Alpes, LIPHY, F-38000 Grenoble, France}
\address{$^3$ CNRS, LIPHY, F-38000 Grenoble, France}
\eads{\mailto{olivier.dauchot@espci.fr}, \mailto{eric.bertin@univ-grenoble-alpes.fr}}
\date{\today}

\begin{abstract} We propose a conservative two-dimensional particle model in
which particles carry a continuous and classical spin. The model includes
standard ferromagnetic interactions between spins of two different particles,
and a nonstandard coupling between spin and velocity of the same particle
inspired by the coupling observed in self-propelled hard discs. Because of
this coupling Galilean invariance is broken and  the conserved linear
momentum associated to translation invariance is not proportional to the
velocity of the center of mass. Also, the dynamics is not invariant under a
global rotation of the spins alone. This, in principle, leaves room for
collective motion and thus raises the question whether collective motion can
arise in Hamiltonian systems. We study the statistical mechanics of such a
system, and show that, in the fully connected (or mean-field) case, a
transition to collective motion does exist in spite of momentum conservation.
Interestingly, the velocity of the center of mass, which in the absence of
Galilean invariance, is a relevant variable, also feeds back on the
magnetization properties, as it acts as an external magnetic field that
smoothens the transition. Molecular dynamics simulations of finite size
systems indeed reveal a rich phase diagram, with a transition from a
disordered to a homogeneous polar phase, but also more complex inhomogeneous
phases with local order interrupted by topological defects.
\end{abstract}
\submitto{\JSTAT}
\maketitle

\section{Introduction}
Spontaneous collective motion, a coordinated motion of an assembly of moving
entities which interact locally without any leader, has recently drawn a lot of
attention from the statistical physics
community~\cite{Ramaswamy:2010bf,Marchetti:2013bp}. Such a phenomenon is present
both in biological systems like motility
assays~\cite{Schaller:2013ej,Schaller:2011dd,schaller467polar,Sumino:2012dw} and
bacterial colonies~\cite{Chen:2012fs,Zhang:2010jn} or on a larger scale insects
swarms~\cite{buhl2006disorder} and flocks of
birds~\cite{ballerini2008interaction,Cavagna:2014vn}, as well as in engineered
systems like driven
colloids~\cite{Palacci:2010hk,Theurkauff:2012ui,Bricard:2013jq,Palacci:2013eu},
droplets~\cite{Thutupalli:2011bv,Izri:2014fva}, or
grains~\cite{Narayan:2007bg,Deseigne:2010gc,Deseigne:2012kn,Kumar:2014wr}. From
a theoretical perspective, such moving individuals are represented as polar
self-propelled particles, that is, particles set in motion by a driving force
directed along the heading vector of the particle. This driving force being
balanced by a friction force, a constant speed is reached in the absence of
interaction with obstacles or other particles. Assuming, as a simplification,
that the speed is always constant leads to minimal models like the Vicsek
model~\cite{vicsek1995novel}, where particles also interact with their neighbors
so as to align their velocity vectors, up to some noise. In other words,
Vicsek-type models can be thought of as
spin-models~\cite{toner1995long,Solon:2013vr}, where particles move along the
direction of their spin instead of remaining fixed on the node of a lattice.
This analogy with spin models is important because in two dimensions (the
dimension in which the Vicsek model is usually defined), the Mermin-Wagner
theorem~\cite{mermin1966absence} prevents the existence of long-range order for
equilibrium models of spins that are invariant under global rotation of the
spins, due to the presence of low-energy excitations called spin-waves~
\cite{Berezinskii70,KT73,KT74}.
The existence of long-range order in the two-dimensional Vicsek model
\cite{vicsek1995novel,toner1995long,Chate:2008is,Gregoire:2004ica,Toner:2012ee}
thus reveals the intrinsic non-equilibrium character of the model, and to some
extent, of the phenomenon of collective motion itself. 
The presence of long-range order in the Vicsek model can be understood  as resulting
from the continuous evolution of the neighborhood of a given particle, which
successively interact with different particles.

As mentioned above, an important simplification of Vicsek-type models is to
identify the direction of motion with that of the heading vector of the
particles. At low enough density, when the relaxation time of the velocity is
short as compared to the typical time between successive interactions, this
approximation is well-justified. In a denser regime however, as observed in
experiments on shaken polar grains~\cite{Deseigne:2010gc,Deseigne:2012kn},
velocity and heading vectors may have different directions, and it is a priori
relevant to consider them as distinct dynamical variables, as done
in~\cite{Weber:2013bj,Lam:2015vh}, assuming a suitable coupling between velocity
and heading. In this situation, we are thus dealing with a fluid of particles
carrying a (classical) spin.

When considering both velocity and heading (or spin) variables, the possible
existence of collective motion at equilibrium cannot be immediately ruled out by
standard arguments if spins and velocities are coupled. The first argument (in
two dimension) is the Mermin--Wagner theorem, but its applicability is not
granted if spins interact with velocities, because the system is then no longer
invariant under a rotation of the spins alone. The second argument is that for
an isolated system at equilibrium the momentum is conserved and no spontaneous
global motion can emerge. Alternatively, boundaries or a substrate may break
momentum conservation, but they act as a momentum sink, also preventing
collective motion. These arguments are however valid only if the standard
relation ${\bf p}=m{\bf v}$ between momentum and velocity holds. Although very
general, it is well-known that such a relation breaks down when the `potential
energy' (the potential term in the Lagrangian) depends on the velocities as
is the case for charges in magnetic field. A third argument related to the
second one, is that at equilibrium the center-of-mass velocity is conjugated, in
a thermodynamic sense, to momentum---just like temperature is conjugated to
energy---and is thus equal to that of the surrounding medium
\cite{diu1989physique}. Invoking Galilean invariance, one then usually sets
the center-of-mass velocity
to zero. However Galilean invariance does not hold in those above situations
where the potential term in the Lagrangian depends on the velocities,
so that the center-of-mass velocity should be a relevant parameter in this
case. A natural question is thus to investigate whether a coupling between
velocity and spin variables could break the Galilean invariance, modify the
standard definition of momentum, and as a result allow for the possibility of
collective motion at equilibrium---equilibrium being understood in the
generic sense of the statistical steady-state of a Hamiltonian system, in the
absence of external forcing.

In this paper, we investigate this issue by considering a simple two-dimensional
model of point-like particles carrying a spin and evolving according to a
conservative dynamics coupling spin and velocity in a minimal way. The dynamics
is originally defined in a Lagrangian formalism, from which a Hamiltonian
formulation is derived. A non-standard expression of the momentum of each
particle in terms of its velocity and spin is obtained. We study the statistical
mechanics of such a system, and show that in the fully connected (or mean-field) case
a transition to collective motion occurs. The velocity of the center of mass,
which in the absence of Galilean invariance, is a relevant variable, also feeds
back on the magnetization properties: it acts as an external magnetic field
that smoothens the transition and stabilizes non trivial local minima of the free energy. 
Molecular dynamics (MD) simulations of finite size systems confirm the existence of a 
homogeneous polar phase. Increasing sizes, the system organizes into domains of collectively 
moving particles structured around topological defects.

\section{The model}
\subsection{Definition}
\label{definition}
We start by considering a liquid of XY-spins in two dimensions. The Lagrangian
of such a model is described by $L = L_\text{r} + L_\text{s}$ with:
\begin{subequations}
\label{eq:lagrangienXY}
\begin{align}
  \label{eq:Lr}
  L_\text{r} &= \sum_{i=1}^N \Bigg( \frac{m}{2} \dot {\bfr}_i^2 - \frac{1}{2}\sum_{k(\neq i)} U(r_{ik}) 
\Bigg)\\
  \label{eq:Ls}
  L_\text{s} &= \sum_{i=1}^N \Bigg( \frac{I}{2} \dot\theta_i^2 +\frac{1}{2} \sum_{k(\neq i)} J(r_{ik}) 
{\bfs_i}\cdot {\bfs_k} + {\bfh}\cdot {\bfs_i}\Bigg)
\end{align}
\end{subequations}
where $r_{ik} \equiv |\bfr_k-\bfr_i|$, with $\bfr_i$, $\theta_i$ and $\bfs_i$,
denoting the position, the angle and the spin of each particle respectively
(the dot on top of a variable denotes time derivative). Note that the spin
$\bfs_i$ is defined as $\bfs_i= {\hat\bfe}(\theta_i) \equiv \cos\theta_i
\hat\bfe_x + \sin\theta_i \hat\bfe_y$. The other parameters appearing in the
Lagrangian are the mass $m$ of the particles, their moment of inertia $I$
associated to spin rotation, the coupling $J=J_0 j(r_{ik})$ between spins, the
external field $\bfh$, and the interaction potential $U(r_{ik})$ between
particles (e.g., hard sphere or Lennard-Jones potential). With the model as it
stands, there is coordination of spins at low temperature. However there
is no coupling between spin and particle motion, so that no collective motion
can emerge. To couple the  motion of the particles to the spin we add to the
Lagrangian a term
\begin{equation}
L_\text{sv} = K \sum_{i=1}^N {\bfs}_i \cdot \dot\bfr_i \,,
\label{eq:Lsv}
\end{equation}
where $K$ is a coupling constant between spin and velocity, which is
expected to favor the alignment of the velocity of particle $i$ to its
\emph{own} spin $\bfs_i$. Apart from the fact that it should contribute to
align velocities when spins align --although we shall see that the effect is
really indirect and counter-intuitive-- this term is motivated by the
observation of such self-alignment in real systems of self-propelled
grains~\cite{Weber:2013bj} and has been identified as a key ingredient for the
dynamics of self-propelled discs~\cite{Lam:2015vh}.

Leaving aside the potential term $U$, although we shall reintroduce it when
moving to the molecular dynamics simulations, our starting point is thus the
following dimensionless Lagrangian:
\begin{equation}
  \label{eq:lagrange}
  L=\sum_{i=1}^N\Bigg(\frac{1}{2} \dot{ \bfr}_i^2 +\frac{1}{2} \dot{\theta}_i^2
  + (K\dot{\bfr}_i \!+\! \bfh) \cdot \bfs_i
  +\frac{1}{2} \sum_{k(\neq i)} j(r_{ik}){\bfs_i}\cdot {\bfs_k}\Bigg),
\end{equation}
where we used the following redefinitions
\begin{subequations}
\begin{align}
  &\bfr/\sqrt{I/m} \rightarrow \bfr,\quad
  t/\sqrt{I/J_0} \rightarrow t,\\
  &\frac{K}{\sqrt{{mJ_0}}} \rightarrow K,\quad
  \frac{\bfh}{J_0} \rightarrow \bfh, \quad L/J_0 \rightarrow L.
\end{align}
\end{subequations}
The parameter $K$ now controls the strength of the alignment between spin and
velocity. The Euler--Lagrange equations then read
\begin{subequations}
\label{eq:lagrangegroupreq}
\begin{align}
  \ddot{\bfr}_i &= -K\dot\theta_i\hat\bfe_{\perp,i}+\sum_{k(\neq i)} \frac{\partial j(r_{ik})}{\partial\bfr_i}
\cos \theta_{ik} \label{eq:lagr}\\
  \ddot\theta_i &=  (K\dot{\bfr}_i + \bfh) \cdot {\hat\bfe}_{\perp,i} +\sum\limits_{k(\neq i)} j(r_{ik})\sin 
\theta_{ik}, \label{eq:lagtheta}
\end{align}
\end{subequations}
where $\theta_{ik}=\theta_k-\theta_i$ and $\hat\bfe_{\perp,i}$ is a unit vector
perpendicular to the spin, defined as $\hat\bfe_{\perp,i} = \hat\bfe(\theta_i+\frac{\pi}{2})$.
Finally, we reformulate the dynamics in the Hamiltonian formalism. The momenta
$\bfp_i$ and $\omega_i$ conjugate to the positions $\bfr_i$ and angles
$\theta_i$ read:
\begin{subequations}
\label{eq:conjmoment}
\begin{align}
  \bfp_i&=\frac{\partial L}{\partial \dot\bfr_i}=\dot\bfr_i +K\bfs_i,\label{eq:conjp}\\
  \omega_i&=\frac{\partial L}{\partial \dot\theta_i}=\dot\theta_i.\label{eq:conjomega}
\end{align}
\end{subequations}
The spin of the particle acts as an internal potential vector and only for $K=0$
is the momentum equal to the velocity. The Hamiltonian is obtained by the usual
Legendre transform of the Lagrangian
\begin{equation}
H = \sum_{i=1}^N \left( \bfp_i \cdot \dot\bfr_i + \omega_i {\dot \theta}_i \right) - L.
\end{equation}
Expressing $\dot\bfr_i$ and $\dot \theta_i$ in terms of the conjugate variables
$\bfp_i$ and $\omega_i$, one finds, discarding an irrelevant constant term:
\begin{equation}
  \label{eq:hamiltonian}
  H =\sum_{i=1}^N \Bigg(\frac{\bfp_i^2}{2}+\frac{\omega_i^2}{2}-(K\bfp_i+ \bfh) \cdot \bfs_i
  -\frac{1}{2}\sum_{k(\neq i)} j(r_{ik}) {\bfs_i}\cdot {\bfs_k} \Bigg).
\end{equation}
The equations of motion follow:
\begin{subequations}
  \label{eq:hamiltongroupreq}
  \begin{align}
    \dot\bfr_i &= \frac{\partial H}{\partial\bfp_i} =\bfp_i-K\bfs_i \label{eq:hamreq}\\
    \dot{\bfp}_i &= -\frac{\partial H}{\partial \bfr_i} =\sum_{k(\neq i)}\frac{\partial j(r_{ik})}{\partial\bfr_i}
\cos \theta_{ik}\label{eq:hampeq}\\
    \dot \theta_i &= \frac{\partial H}{\partial\omega_i} = \omega_i\label{eq:hamtheeq}\\
    \dot\omega_i &= -\frac{\partial H}{\partial\theta_i} = (K\bfp_i+\bfh) \cdot \hat\bfe_{\perp,i}+
\sum_{k(\neq i)} j(r_{ik})\sin \theta_{ik}\label{eq:hamomeeq}.
  \end{align}
\end{subequations}
Let us note that, up to a constant the Hamiltonian can be formally written as
\begin{equation}
  \label{eq:hamiltonian2}
  H = \sum_{i=1}^N \Bigg(\frac{\dot\bfr_i^2}{2}+\frac{\omega_i^2}{2}- \bfh \cdot \bfs_i
     -\frac{1}{2}\sum_{k(\neq i)} j(r_{ik}) {\bfs_i}\cdot {\bfs_k} \Bigg).
\end{equation}
Although this is not a proper formulation in terms of the canonical variables,
it shows that written in terms of kinetic and potential energy, the Hamiltonian
takes a form similar to the one of a standard liquid of particles
carrying an XY-spin. The subtlety hidden in this misleadingly simple
formulation is that $\dot\bfr_i$ is not proportional to $\bfp_i$.

\subsection{Continuous Symmetries}%
\label{sec:symm}
Taking advantage of the Lagrangian formulation, we apply Noether's theorem to
obtain the conserved quantities of the dynamics, associated to the continuous
transformation under which the Lagrangian is invariant.

First of all, the Lagrangian is invariant under time and space translation, so
that the dynamics conserves the energy $E=H$ and the \emph{generalized} linear
momentum $\bfP$ obtained by summing \eqref{eq:conjp} over all particles:
\begin{equation} \label{eq:PVGKS}
  \bfP =N\bfV+K\bfM,
\end{equation}
where $\bfV$ is the center of mass velocity and $\bfM=\sum_{i=1}^N \bfs_i$ is the
total magnetization of the system. This simple but essential formula relates the
total momentum of the system to the collective motion, described by the velocity
of the center of mass, and the total magnetization of the system. Although the
generalized linear momentum considered here is not proportional to the velocity
of the center of mass, this conservation law is in stark contrast with most
two-dimensional models for collective motion, for which momentum is not
conserved.

The equations of motion are not invariant under a rotation of the system (i.e.,
a global rotation of the position of particles), as can be seen from
\eqref{eq:lagrange}, where the scalar product is not invariant, as the spins do
not rotate, and also because of the presence of the external field $\bfh$.%
\footnote{For $\bfh=\bfzero$, the invariance is however recovered under a
transformation in which the spins are rotated together with the positions of
the particles.  In that case a supplementary conserved quantity is  $\mathcal L
= L_z + \sum_{i=1}^N\omega_i $ with  $L_z = \sum_{i=1}^N \hat\bfe_z \cdot
(\bfr_i \times \bfp_i)$  the usual angular momentum. In practice, even
when the field $\bfh=0$, this rotational invariance is broken because of
boundary conditions.}

Finally, let us stress a significant difference with the standard
XY-model~\cite{Berezinskii70,KT73,KT74}. In the present case, there is no
symmetry under rotation of the spins alone. Accordingly spin waves are not slow
modes of the dynamics and the Mermin-Wagner theorem (at least in its
form) does not apply, opening the way for possible long-range order.

\subsection{Time reversibility}
Interestingly the dynamics is not time reversible. Applying the transformation
$t\rightarrow-t$:
\begin{subequations}
\begin{align}
  \bfr_i &\rightarrow \bfr_i,\quad  \dot\bfr_i \rightarrow -\dot\bfr_i,\quad \ddot{\bfr}_i \rightarrow 
\ddot{\bfr}_i,\\
  \theta_i &\rightarrow \theta_i,\quad \dot{\theta}_i \rightarrow - \dot{\theta}_i,\quad  \ddot{\theta}_i 
\rightarrow \ddot{\theta}_i
\end{align}
\end{subequations}
to the Euler-Lagrange equations~\eqref{eq:lagrangegroupreq}, they transform into
\begin{subequations}
\begin{align}
  \ddot{\bfr}_i &= K\dot\theta_i\hat\bfe_{\perp,i} + \sum_{k(\neq i)}\frac{\partial j(r_{ik)}}{\partial\bfr_i}
\cos \theta_{ik}\\
  \ddot\theta_i &= (-K\dot\bfr_i+ \bfh) \cdot \hat\bfe_{\perp,i} + \sum_{k(\neq i)} j(r_{ik})\sin \theta_{ik},
\end{align}
\end{subequations}
which differ from the original ones [Eq.~\eqref{eq:lagrangegroupreq}], unless
$K=0$. Note that although the Hamiltonian is symmetric under time reversal, it
does not imply time reversibility of the trajectories because $\bfp_i$
transforms under time reversal into $\bfp_i' = - \bfp_i + 2K \bfs_i$, as can
also be checked directly from Eq.~\eqref{eq:hamiltongroupreq}.

The invariance of the trajectories can be recovered using the more general
transformation $t \to -t$ and $\bfs_i \to -\bfs_i$. In this case, one has
$\bfp_i \to - \bfp_i$ and both the Hamiltonian and the trajectories are
invariant. The invariance under this more general transformation can be
interpreted as a generalized form of microreversibility. At a heuristic
level, this suggests that a generalized form of detailed balance, associated to
a uniform measure in the micro-canonical ensemble, may hold.

\subsection{Galilean Invariance}
Also central in classical mechanics is the Galilean invariance, which states
that the equations of motion are identical in different coordinate systems
moving with constant velocity with respect to each other. Applying a Galilean
transformation to the Euler-Lagrange equations~\eqref{eq:lagrangegroupreq}
\begin{align}
  &\bfr_i'=\bfr-\bfV_0 t, &\dot\bfr_i'=\dot\bfr_i-\bfV_0, \quad &\ddot{\bfr}'_i=\ddot {\bfr}_i,
\end{align}
while keeping $\theta_i$ and its derivatives unchanged, one finds in the new
frame, dropping primes to lighten notations,
\begin{align}
  \ddot{\bfr}_i &= K\dot\theta_i\hat\bfe_{\perp,i}+\sum_{k(\neq i)} \frac{\partial j(r_{ik})}{\partial\bfr_i}
\cos \theta_{ik}\\
  \ddot\theta_i &= K\bfV_0 \cdot \hat\bfe_{\perp,i} + (K\dot\bfr_i+ \bfh) \cdot \hat\bfe_{\perp,i} + 
\sum_{k(\neq i)} j(r_{ik})\sin \theta_{ik},
\end{align}
where the extra term $K\bfV_0 \cdot \hat\bfe_{\perp,i}$ in the last equation breaks
Galilean invariance.

This absence of Galilean invariance is actually connected to the fact that the
total momentum $\bfP$ is not proportional to the velocity $\bfV$ of the center
of mass, as can be seen from the following argument. The term responsible for
the broken Galilean invariance is the term $L_\text{sv}$ coupling spin and
velocity [Eq.~\eqref{eq:Lsv}]. Let us momentarily consider a general coupling
term $L_\text{sv}$, assuming simply that it depends only on the spins ${\bfs}_i$
and the velocities $\dot\bfr_i$, $i=1,\dots, N$. Then the total momentum reads
\begin{equation}
  \bfP = N \bfV + \sum_{i=1}^N \frac{\partial L_\text{sv}}{\partial \dot\bfr_i}
\end{equation}
and is thus generically not proportional to the velocity of the center of mass.
However, if the coupling term $L_\text{sv}$ satisfies Galilean invariance, one
has $L_\text{sv}(\dot\bfr_i-\delta \bfV_0, {\bfs}_i) = L_\text{sv}(\dot\bfr_i, {\bfs}_i)$,
where $\delta \bfV_0$ is the velocity shift associated to an infinitesimal
Galilean transformation, resulting in
\begin{equation}
  \sum_{i=1}^N \frac{\partial L_\text{sv}}{\partial \dot\bfr_i} = 0
\end{equation}
so that one recovers $\bfP = N \bfV$. In the following, when discussing the role
of the broken Galilean invariance, we may thus consider the following coupling term,
\begin{equation} \label{eq:coupling-eta}
  L_\text{sv} = \sum_{i=1}^N K \Bigl( {\bfs}_i - \frac{\eta}{N}\sum_{k=1}^N {\bfs}_k \Bigr) \cdot \dot
\bfr_i .
\end{equation}
For $\eta=0$, this term corresponds to the coupling defined in
Eq.~\eqref{eq:Lsv}, while for $\eta=1$, Galilean invariance is recovered and
$\bfP = N \bfV$.

\subsection{Motion of a single particle}
\begin{figure}[b]
  \centering%
  \includegraphics{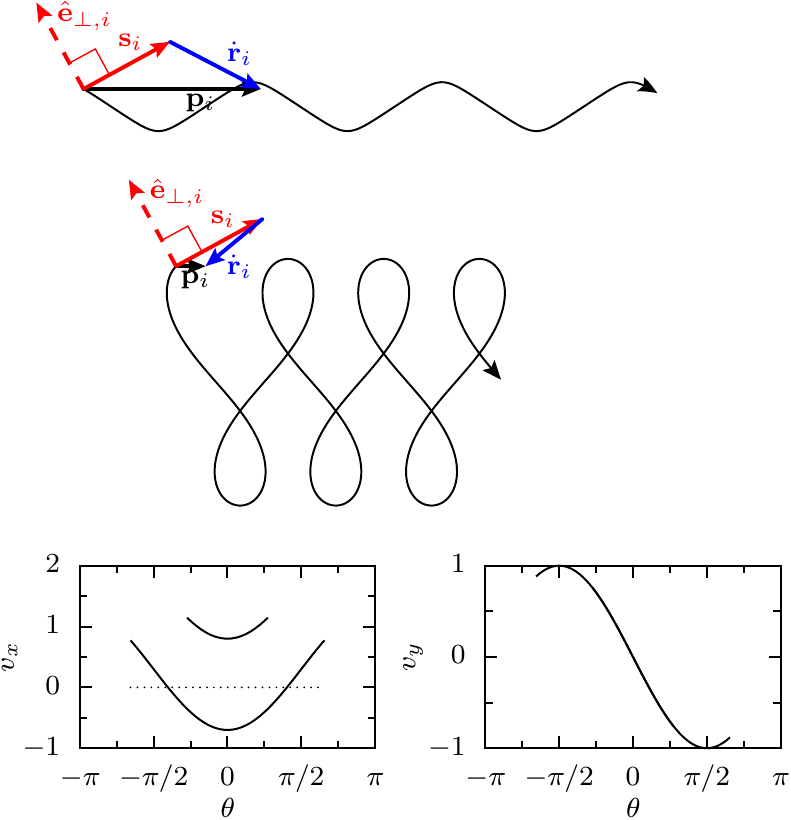}%
  \caption{Dynamics of a single spin in the cases $Kp_0>1$ (top) and
  $Kp_0<1$ (middle). The momentum $\bfp_i = \bfp_0$ is conserved, the spin
  $\bfs_i$ oscillates around the direction given by $\bfp_i$ with a frequency
  $\sqrt{K p_0}$. The velocity $\dot\bfr_i$ is then entirely set by
  Eq.~\eqref{eq:hamreq}. The plots in the bottom row show that only the
  $x$-components of the trajectories differ, whereas the $y$-components
  oscillate on the same trajectory.}%
  \label{fig:DirSpin}
\end{figure}

To gain some intuition on particle motion, it is useful to analyze the equations
of motion for a single and free particle $(j=0, \bfh=0)$. The individual
momentum is conserved, $\bfp(t)=\bfp_0$. Taking the temporal derivative of
Eq.~\eqref{eq:hamtheeq} and using Eq.~\eqref{eq:hamomeeq}, one gets
\begin{equation}
  \ddot\theta = K\bfp \cdot \hat\bfe_{\perp} = -K p_0\sin\theta,
\label{eqapp:dt2}
\end{equation}
where the coordinate system has been set such that $\hat\bfe_x = \hat\bfp_0/p_0 $, and
$\bfp \cdot \bfs = p_0 \cos\theta$. One recognizes the equation of motion of a
pendulum: the spin $\bfs$ oscillates around the direction $\hat\bfp_0$ with a
frequency $\sqrt{K p_0}$. The velocity of the particle follows from
Eq.~\eqref{eq:hamreq}, as illustrated on Fig.~\ref{fig:DirSpin} in the case
$K=1$.
Using the conservation of the angular momentum $\mathcal L = -y(t) p_0 +
\dot\theta$ (see footnote 1), one finds $y(t) = y(0) + ({\dot \theta}(t) - {\dot
\theta(0)})/p_0$. The motion of the particle, perpendicular to the direction of
its momentum, follows the periodic motion of $\dot\theta$. The dynamics along
the direction of the momentum is more complicated. From Eq.~\eqref{eq:conjp} one
has $\dot x = p_0 - K \cos\theta$. For small enough $K$, $K/p_0 <1$, the
particle always moves in the direction of its momentum. However, for larger K,
the motion strongly depends on the dynamics of $\theta$; and, for small
$\theta$, that is when the spin points to a direction close to that of the
momentum, the particle moves into the opposite direction! This counter-intuitive
behavior is deeply rooted in the conservation of $\bfp = \dot\bfr_i + K \bfs$. In the
following we shall see that the constraints imposed on the dynamics by the
conservation laws, together with the coupling between the spins and the
velocities, lead at the mean-field level to the onset of collective motion.

\section{Statistical description} \label{sect-stat}
\subsection{Distribution of microscopic configurations}
Starting from the Hamiltonian formulation of the model a Liouville equation can
be written for the probability density function of the phase-space point
$\mathcal{C}=({\bfr}^N, \theta^N, {\bfp}^N, \omega^N)$. It follows that at
equilibrium the probability distribution is a function of the conserved
quantities of the dynamics, namely the energy $E$ and the linear momentum $\bfP$.
In the microcanonical ensemble, the conserved quantities cannot be exchanged
with the environment; they have constant values $E_0$ and ${\bfP}_0$
respectively, and the distribution $\mathcal{P}(\mathcal C)$ is given by
\begin{equation} \label{eq-distmicrocan}
  \mathcal{P}_\text{mc}(\mathcal C) = \frac{1}{\Omega} \, \delta(E-E_0)\delta(\bfP-\bfP_0),
\end{equation}
where $\Omega$ is the microcanonical partition function, defined by normalizing
$\mathcal{P}_\text{mc}(\mathcal C)$ to $1$. In other words, all configurations
with the same values of the conserved quantities are equally probable.

In practice, it is more convenient from a computational viewpoint to work in the
canonical ensemble where the conserved quantities are exchanged with a
reservoir. The probability measure in the canonical ensemble is then given by
\begin{equation}
\label{eq-distcan0}
  \mathcal{P}(\mathcal C) = \frac{1}{Z(\beta,\bfalpha)} e^{-\beta(E-\bfalpha \cdot \bfP)},
\end{equation}
where $\beta$ and $\bfalpha$ are intensive control parameters of the
reservoir that determine the thermal averages of $E$ and $\bfP$ ($\beta$
denotes, as usual, the inverse temperature).

\subsection{Computation of average moments}
The distribution \eqref{eq-distcan0} of the microscopic configurations can be
written more explicitly as
\begin{equation}
  \label{eq:dist-can}
  P(\bfr^N,\theta^N,\bfp^N,\omega^N)
  = \frac{1}{Z(\beta,\bfalpha)} \prod_{i=1}^{N} \phi(\bfp_i, \theta_i, \omega_i)
  \prod_{i<k} \psi(r_{ik}, \theta_{ik})
\end{equation}
with
\begin{align} \nonumber
  \phi(\bfp_i, \theta_i, \omega_i)
   &= \exp \Bigl[ -\beta \Bigl( \frac{ {\bfp}_i^2}{2} + \frac{\omega_i^2}{2} - \bfh \cdot \bfs_i - 
(\bfalpha{+}K{\bfs}_i)\cdot {\bfp}_i \Bigr)\Bigr]\\
  \psi(r_{ik}, \theta_{ik})
   &= \exp\Bigl(\beta j(r_{ik}) \cos\theta_{ik}\Bigr) \, ,
\end{align}
and where
\begin{equation}
  \label{eq:Z}
  Z(\beta,\bfalpha) = Z_\omega \int d\theta^N Z_p(\theta^N)\,\, Z_r(\theta^N)
\end{equation}
is the partition function, with
\begin{align}
  \label{eq:Zw}
  Z_\omega &= \prod_{i=1}^{N}\int d\omega_i \, e^{-\beta \frac{\omega_i^2}{2}}\\
  \label{eq:Zp}
  Z_p (\theta^N) &= \prod_{i=1}^{N} \int d\bfp_i \, e^{-\beta \bigl( \frac{ {\bfp}_i^2}{2} - (\bfalpha + 
K{\bfs}_i)\cdot {\bfp}_i - \bfh \cdot \bfs_i \bigr)} \\
  \label{eq:Zr}
  Z_r (\theta^N) &= \int d\bfr^N  \prod_{i<k} e^{\beta j(r_{ik}) \cos\theta_{ik}}   .
\end{align}
From the distribution \eqref{eq:dist-can}, one can easily compute the
first and second moments of the momenta $\bfp_i$ and $\omega_i$. For the first
moments, one finds $\langle \omega_i \rangle = 0$ and
\begin{equation}
  \label{eq:Avgp}
  \langle {\bfp}_i \rangle = \bfalpha + K \langle \bfs_i \rangle.
\end{equation}
Using Eq.~\eqref{eq:conjp} in Eq.~\eqref{eq:Avgp} yields $\bfalpha
=\langle\dot\bfr_i\rangle$. Taking the sum over all particles and dividing by
$N$ then leads to
\begin{equation}
  \label{eq:alpha}
  \bfalpha = \langle\bfV\rangle.
\end{equation}
This relation, which shows that the intensive parameter $\bfalpha$
associated to the conservation of momentum is the averaged velocity of the
center of mass is not specific to the present model. It is a general property
of equilibrium systems~\cite{diu1989physique}. However, in the presence of
Galilean invariance one can arbitrarily set $\langle\bfV\rangle=0$ and safely
ignore $\bfalpha$ in the partition function. Here because of the broken
Galilean invariance we shall on the contrary keep it as a true and independent
intensive thermodynamic parameter. Inserting Eq.~\eqref{eq:alpha} into
Eq.~\eqref{eq:Avgp}, we end up with
\begin{equation}
  \langle {\bfp}_i \rangle = \langle\bfV\rangle + K \langle \bfs_i \rangle,
\end{equation}
which can be seen as a `local' counterpart of Eq.~\eqref{eq:PVGKS}.

For the second moment, one finds a generalization of the equipartition
relations, with $kT = \beta^{-1}$,
\begin{align}
  \label{eq:kbT}
  \langle \omega_i^2 \rangle &= kT, \\
  \left(\langle {\bfp}_i^2 \rangle - \langle {\bfp}_i \rangle^2\right)  - K^2
  \left(\langle {\bfs}_i^2 \rangle - \langle {\bfs}_i \rangle^2\right) &= 2kT.
\end{align}
Hence not only are the average values of momentum and spin related, as could
have been anticipated from Eq.~\eqref{eq:PVGKS}, but so are also their
fluctuations. Note however that the temperature is \emph{not} proportional to
the total kinetic energy because of the coupling between the translational
velocities and the spins.

%
\subsection{Phase transition in the fully connected model}
\label{sec:fc}
We now study the behavior of the mean magnetization as a function of
temperature and center of mass velocity (or equivalently, $\bfalpha$). To keep
the discussion at a simple enough level, we focus on the fully connected
geometry, which is akin to a mean-field approximation (note that the
momenta ${\bfp}_i$ and spins ${\bfs}_i$ however remain two-dimensional
vectors). In this case, the interaction amplitude $j(|\bfr_i - \bfr_k|)$ is
simply a constant, independent of the distance between particles. To ensure
that energy remains extensive, we take this constant to be equal to $1/N$. The
interaction term can then be rewritten as follows
\begin{align} \nonumber
  \frac{1}{N} \sum_{i=1}^N  \sum_{k(\neq i)} {\bfs_i} \cdot {\bfs_k}
    &= \frac{1}{N} \sum_{i=1}^N \bfs_i \cdot \Bigl(\sum_{k=1}^N  \bfs_k -\bfs_i\Bigr)\\
    \label{s_doublesum}
    &= N\bfm^2-1,
\end{align}
where $\bfm(\theta^N) = N^{-1} \sum_{i=1}^N \bfs_i$ is the magnetization per
spin. Disregarding the constant term (which amounts to a shift in the energy
reference), the integrations of Eqs.~\eqref{eq:Zw}, \eqref{eq:Zp} and
\eqref{eq:Zr} over $\omega^N$, $\bfr^N$ and $\bfp^N$ are readily computed,
yielding
\begin{equation} \label{eq:Zcm2}
  Z(\beta,\bfalpha,N,V) = \Bigl(\frac {2\pi}{\beta}\Bigr)^{\frac{3N}{2}}\, V^N\, e^{\frac{\beta}{2} N(K^2 
+ \bfalpha^2)}\, Z_\theta
\end{equation}
where $V$ is the volume occupied by the system. The integral that remains to be
computed is
\begin{equation}\label{eq:Ztheta}
  Z_\theta = \int d\theta^{N} \, e^{\frac{1}{2}N\beta\bfm^2 + N\beta\bfm\cdot(K\bfalpha+\bfh)},
\end{equation}
where one recognizes the mean field partition function of the conventional
XY-model in the presence of an external field $\bfh_\text{eff} = K \bfalpha +
\bfh$~\cite{chaikin2000principles}. Following standard techniques (see
Appendix~A), one finds
\begin{equation} \label{eq:Ztheta3}
  Z_\theta = \frac{N\beta}{2\pi}\int du_1 du_2 \, e^{- N\beta \calF(\bfu)},
\end{equation}
where the function $\calF(\bfu)$ is given by
\begin{equation} \label{eq:def-Fu}
  \calF(\bfu) = \frac{\bfu^2}{2}
  -\frac{1}{\beta} \ln \bigl[2\pi I_0 \bigl(\beta\gamma(\bfu)\bigr)\bigr]
\end{equation}
with $I_n$, the modified Bessel function of order $n$ and $\gamma(\bfu) = \left|
\bfgamma(\bfu) \right| = \left|K \bfalpha +  \bfh + \bfu \right|$. In the
large $N$ limit, the integral in Eq.~\eqref{eq:Ztheta3} can be computed using
the saddle point approximation, yielding to exponential order
\begin{equation} \label{eq:Ztheta4}
  Z_\theta \sim e^{-N \beta \calF(\bfu^*)}
\end{equation}
where the saddle-point $\bfu^*(\bfalpha,\beta,\bfh)$ is given by
\begin{equation}\label{eq:MeanFieldEq}
  \frac{\partial\calF}{\partial \bfu}(\bfu^*) = \bfu^*
- \frac{I_1(\beta\gamma(\bfu^*))}{I_0(\beta\gamma(\bfu^*))} \, \hat{\bfgamma}(\bfu^*)= 0
\end{equation}
with $\hat{\bfgamma}(\bfu^*)$ the unit vector
$\bfgamma(\bfu^*)/\gamma(\bfu^*)$. Using Eq.~\eqref{eq:Zcm2} and Stirling's
approximation, the free energy density $f$ then reads
\begin{multline}
  f(\beta,\bfalpha,N/V)
   := -\frac{kT}{N} \ln\frac{Z(\beta,\bfalpha,N,V)}{N!} \\
   \sim\calF(\bfu^*) - \frac{1}{\beta}\Bigl(\frac{3}{2}\ln\frac{2\pi}{\beta}-\ln\frac{N}{V} + 1\Bigr)
    - \frac{\bfalpha^2}{2}.
\end{multline}
It is a function of intensive variables only. Finally the average magnetization
$\langle \bfm \rangle$ per particle is:
\begin{equation}
  \langle \bfm \rangle = -\frac{\partial f}{\partial \bfh}
  = -\frac{\partial \bfu^*}{\partial \bfh}\frac{\partial\calF}{\partial \bfu}(\bfu^*)
    -\frac{\partial\calF}{\partial \bfh}(\bfu^*).
\end{equation}
Since $\partial\calF/\partial \bfu(\bfu^*)=0$, we end up with $\langle \bfm
\rangle = -\frac{\partial\calF}{\partial \bfh}(\bfu^*)$. Combining this last
result with the definition of $\bfgamma$ and Eqs.~\eqref{eq:def-Fu} and
\eqref{eq:MeanFieldEq}, we get that $\langle \bfm \rangle = \bfu^*$. Hence
$\langle \bfm \rangle$ can be obtained from the self-consistent equation
\eqref{eq:MeanFieldEq}, simply replacing $\bfu^*$ by $\langle \bfm \rangle$.%
\begin{figure}[t!]
  \centering
  \includegraphics{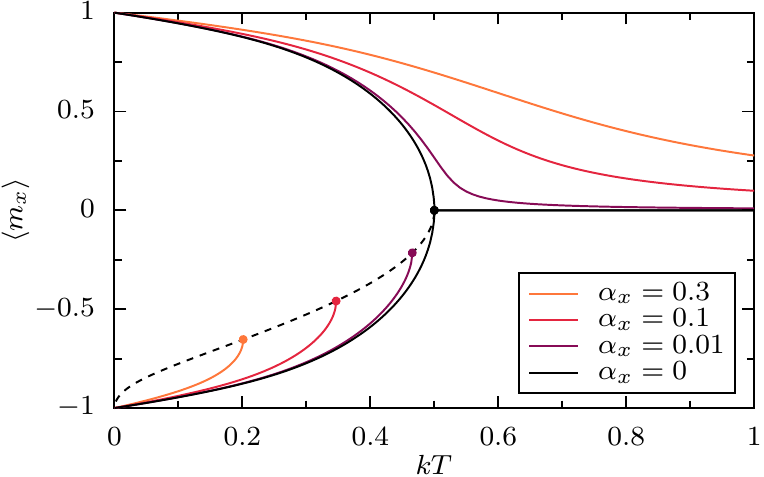}%
  \caption{Magnetization of the fully connected model, in the absence of
  external field $(\bfh=\bfzero)$, but with non-zero effective field
  $\bfh_\text{eff}=K\bfalpha$. Solid curves are obtained from numerically
  minimizing $\calF(\bfu)$. Parameters are $K=1$, $\alpha_y=0$.}%
  \label{fig:MeanFieldAnal}
\end{figure}

Using $\frac{\partial \ln Z}{\partial \beta} = - \langle H \rangle +
\bfalpha\cdot\langle\bfP\rangle$, we also obtain the average energy per
particle $\langle e \rangle = \langle H \rangle/N$:
\begin{equation}
  \label{fc:e}
  \langle e \rangle = \frac{3}{2\beta} + \frac{\bfalpha^2 - {\bfu^*}^2}{2} - \bfh \cdot \bfu^*.
\end{equation}
Eq.~\eqref{eq:MeanFieldEq} can be solved numerically, the result being
depicted in Fig.~\ref{fig:MeanFieldAnal} for the case $K=1$ in the absence of
external field $(\bfh=\bfzero)$. Without loss of generality we choose
$\bfalpha$ in the $x$-direction. When $\bfalpha=\bfzero$, there is a phase
transition from an isotropic to a magnetized phase at $kT_c=1/2$. For nonzero
values of~$\bfalpha$, we find in the upper half of the figure smooth
magnetization curves which demonstrate that the coupling between the spins and
the particle velocities is encompassed into an effective
field~$\bfh_\text{eff}=\bfh+K\bfalpha$: as long as $K>0$, the average velocity
of the center of mass acts as an external field, polarizing the spins. This
nontrivial effect of the center-of-mass velocity is directly related to the
loss of Galilean invariance. Actually, starting from the more general coupling
term given in Eq.~\eqref{eq:coupling-eta}, one finds that the effective
external field $\bfh_\text{eff}$ is changed into\footnote{To be more specific,
the function $\calF(u)$ is also changed into
\begin{equation*}
  \calF(\bfu) = c\frac{\bfu^2}{2}
-\frac{1}{\beta} \ln \bigl[2\pi I_0\bigl(\beta\gamma(\bfu)\bigr)\bigr]
\end{equation*}
with $c=1+\eta(2-\eta)K^2$ and $\gamma(\bfu)=(1-\eta)K {\bfalpha} + {\bfh} + c{\bfu}$.}
\begin{equation}
  \bfh_\text{eff} = \bfh + K(1-\eta) \bfalpha,
\end{equation}
so that the contribution from $\bfalpha$ to the effective field vanishes for
$\eta=1$, when Galilean invariance is recovered.

In its lower half, Figure~\ref{fig:MeanFieldAnal} shows further solutions.
These are also obtained from minimizing~$\calF(\bfu)$, but here the minima are
local ones. In these solutions, the magnetization~$\bfM$ is \emph{opposite} to
the center-of-mass velocity~$\bfalpha$. We can already anticipate that
these local minima are essential to the low temperature physics of the model
in the microcanonical ensemble: for a system with vanishing
fixed total momentum, Eq.~(\ref{eq:PVGKS}) imposes $\langle \bfm
\rangle = -K\bfalpha$ and the system will select the minima for which
$\bfalpha$ and $\langle \bfm \rangle$ are anti-aligned.

Although observing an ordering transition in a mean-field framework usually
does not come as a surprise, let us emphasize that the present transition is
non-standard even at mean-field level, in the sense that collective motion
cannot be observed in equilibrium systems where momentum is either conserved or
exchanged with a substrate, as long as the relation $\bfP=N\bfV$ is valid. The
transition clearly relies on the coupling between spin and velocity, and
disappears for $K=0$.

Also, the above results only hold for the
fully-connected model, that is at the mean-field level. In the case of the
standard two-dimensional XY-model, it is very well known that mean-field
approximations erroneously predict a transition towards a true long-range
ordered phase at low temperature, which in two dimensions is replaced by the
celebrated Berezinskii--Kosterlitz--Thouless transition towards a
quasi-long-range ordered phase, with zero magnetization, but infinite
correlation length of its fluctuations~\cite{Berezinskii70,KT73,KT74}.
Physically, long-range order in the two-dimensional XY-model is destroyed by
the low-energy spin-wave excitations associated with the invariance of the
dynamics under a continuous rotation of the spins. In the present case, we have
seen in section~\ref{sec:symm} that this symmetry is absent for $K\ne0$
when globally rotating the spins alone. One can however argue that a more
general transformation, rotating both the spins and the velocities, should be
applied to restore the Mermin-Wagner theorem. When a reservoir imposes
$\bfalpha$, or when the conserved momentum has a fixed value $\bfP_0 \neq 0$,
isotropy is actually broken. Yet, when $\bfP_0 =0$, the system is isotropic and
long wave-length excitations involving spins and velocities may be expected to
destroy long-range order.

\section{Molecular Dynamics simulations}

In order to perform numerical simulations of the dynamics described by the
equations of motion~\eqref{eq:hamiltongroupreq}, one needs to specify the
spatial dependence of the ferromagnetic interaction. Doing so, one notices that
the interaction not only leads to alignment of the spins but also to
attraction/repulsion of interacting spins, depending on their relative
orientations. Typically, spins of similar direction cluster together while
oppositely pointed spins repel each other. An aligned phase is thus
expected to end up with all particles being very close together. In order to
avoid this undesired behavior, we add a repulsive potential $U(r_{ik})$ as
introduced in Eq.~\eqref{eq:Lr}, which translates into a repulsive force term
in Eq.~\eqref{eq:hampeq}. In the following, we set
\begin{equation}
  j(r)=(1-r)^2 \quad\text{and}\quad  U(r)= 4(1-r)^4
\end{equation}
for $r\leq1$ and $j(r) = U(r) = 0$ for $r\geq 1$. The interaction range
is thus one unit length and the two interactions are of the same magnitude for
$r=1/2$.
\begin{figure}
  \centering
  \includegraphics{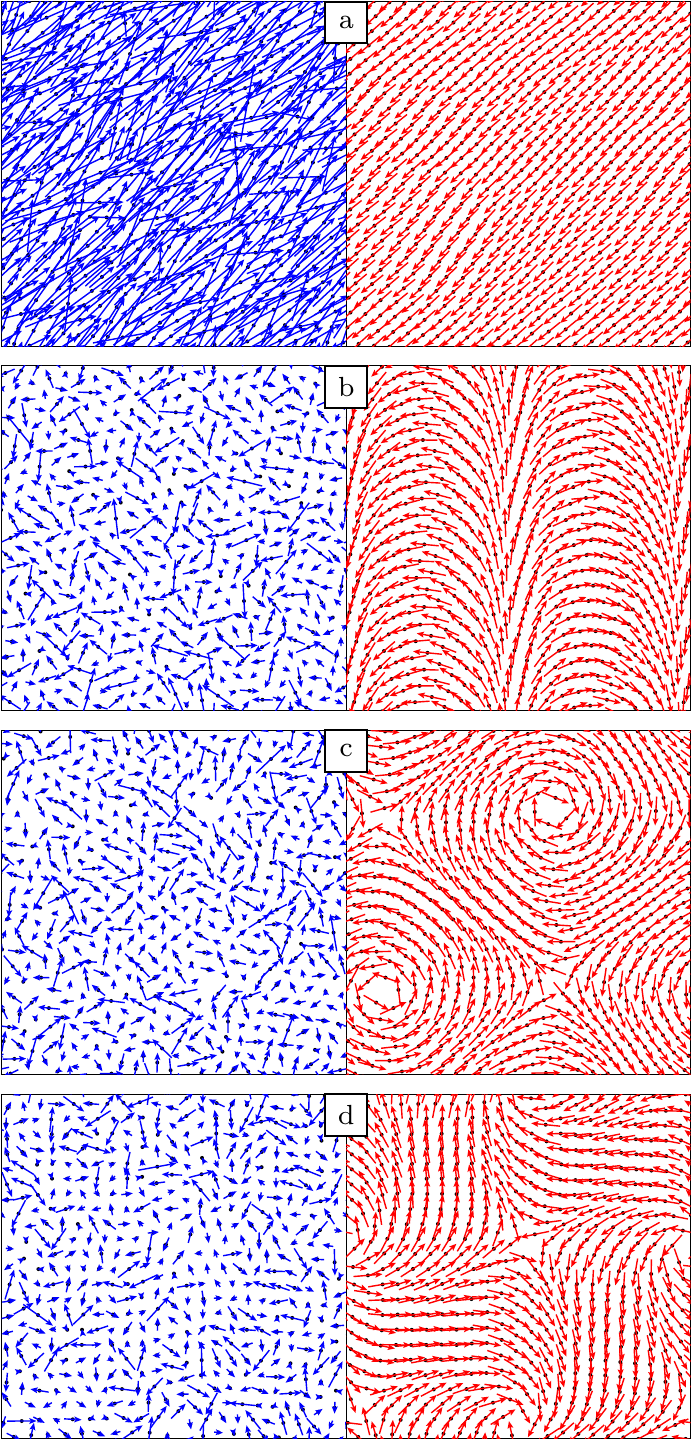}%
  \caption{(a)~Ground state for $K=0.1$. (b--d) Candidates for ground states
  for $K=0.3,0.5,1.0$. Black dots indicate particle positions in the
  $x,y$-plane; left:~blue arrows are~$\dot\bfr_i$; right: red arrows
  are~$\bfs_i$. $N=500, \rho=3.55, \zeta=0.9999$.}%
  \label{fig:gs}
\end{figure}
We perform simulations in the microcanonical ensemble in square boxes of size
$L\times L$, with periodic boundary conditions and a fixed density of particles
$\rho = N/L^2 = 3.55$; $N=[256, 500, 1000]$. The equations of motion are
integrated using a standard fourth-order Runge--Kutta approximation. The total
energy $E_0$ and momentum $\bfP_0$ are set by the initial condition. In the
following, we shall always start from initial conditions where all positions
and spins orientations are random, and all velocities, both $\dot\bfr_i$ and
$\omega_i$, are zero. Hence in the initial conditions the total magnetization
$\bfM_0 =\bfzero$ and the velocity of the center of mass $\bfV_0= \bfzero$, so
that the total momentum $\bfP_0 =\bfzero$. Typically, starting from such an
initial condition, the system reaches a disordered steady state. However,
from the phase transition diagram in Fig.~\ref{fig:MeanFieldAnal}, one
would expect that a system prepared with low enough energy will also be at low
temperature and will thus evolve towards an ordered state with nonzero
magnetization~$\langle \bfm \rangle$. If this were the case, because
$\bfP=\bfzero$ is conserved, the system would acquire a finite velocity of its
center-of-mass $\bfV=-K\langle \bfm \rangle$: \emph{collective motion would set
in spontaneously}.

We first look for the ground state for different values of~$K$. We perform
slow annealing of the system by removing rotational kinetic energy at a
constant rate: every $100$~integration steps, a factor $\zeta<1$ is applied
to all~$\omega_i$. We checked that the annealing rate was small enough to
ensure that we observed (statistically) the same result for different rates.
We also checked that the total momentum~$\bfP$ remained null during the
annealing procedure. Figure~\ref{fig:gs} displays the final states obtained
from this annealing procedure in a system of $N=500$ particles. One observes
quite a rich phase behavior: for $K=0.1$ all spins are indeed aligned as in
panel~(a) and, as anticipated, all velocities align in the opposite
direction, so that collective motion is present. For $K=0.3$, $0.5$, $0.7$,
and $1$ we observe three kinds of different final states occurring at random,
which are depicted in panels (b)~to (d). They have large-scale structure in
their magnetization field, such that the total magnetization vanishes; as a
result $\bfV=\bfzero$: the velocities fluctuate independently, and no global
motion takes place. The three states in Fig.\ref{fig:gs}b--d can be seen as
candidates for the ground state, representing local minima in a free-energy
landscape. The frequency at which one of these states is picked by the
annealing process depends on~$K$ and on the number of particles~$N$. For
$N=500$, we could only find the magnetized ground state for $K=0.1$.

\begin{figure}
  \centering
  \includegraphics{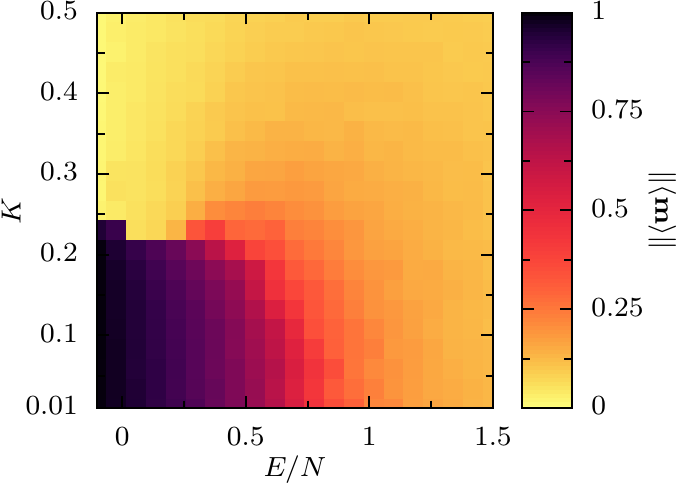}%
  \caption{Parameter-dependence of the average magnetization in the steady
  state as a function of the prescribed energy $E$. Values of $K$ run from
  $0.01$ to $0.5$ in equal steps. In all simulations $\bfP=\bfzero$, $N=256$,
  $\zeta=0.999$, $\rho=3.55$.}%
  \label{fig:KEscan}
\end{figure}

We now investigate how magnetization resists thermal fluctuations.
Figure~\ref{fig:KEscan} displays the phase diagram for systems with zero total
momentum $\bfP =\bfzero$. To obtain it, we slowly anneal the system from a
disordered initial state with energy~$E_0$, but stop the annealing at a
beforehand chosen value of~$E$, which is then conserved. We then wait for the
system to relax and start to average the magnetization. One can see that for
sufficiently small $K$ magnetization survives on a finite range of energy.%
\begin{figure}
  \centering
  \includegraphics{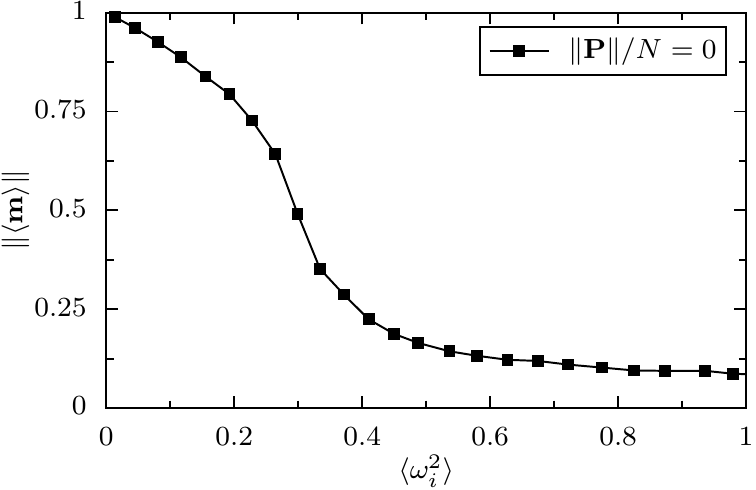}%
  \caption{Average magnetization in microcanonical simulations, as a
  function of the measured temperature. Colors/linestyles correspond to
  different momenta, symbols indicate system sizes: $N=256$ (square), $500$
  (circle), $1000$ (triangle). In all simulations $K=0.1$, $\zeta=0.99$,
  $\rho=3.55$.}%
  \label{fig:compare_fc}
\end{figure}%
\begin{figure}
  \centering
  \includegraphics{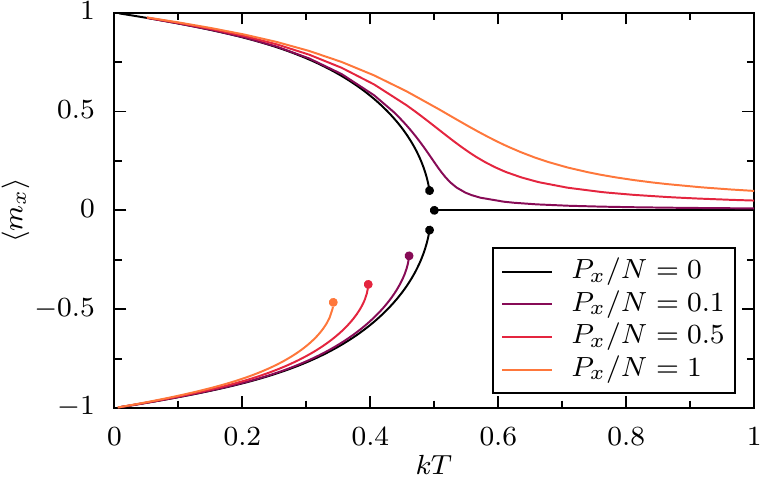}%
  \caption{Curves of constant $\bfP$ in the fully connected model.
  $K=0.1$, $\bfh=\bfzero$, $\alpha_y=0$, $P_y=0$.}%
  \label{fig:constp}
\end{figure}%

Focusing on the case $K=0.1$, for which collective motion is indeed observed
(at least in the finite size system), we investigate the transition towards
the disordered state at high energy. The simulation data is plotted in
Fig.~\ref{fig:compare_fc} for different system sizes and different momenta.
Here, for the purpose of comparison with the canonical case, we choose to
plot the magnetization as a function of $kT$, where the temperature is
measured using the rotational velocity fluctuations, as prescribed by
Eq.~\eqref{eq:kbT}. A crossover from low to high values of the magnetization
is clearly visible when temperature is decreased. To better compare it with
the canonical calculation of Sec.~\ref{sec:fc} above, although no
quantitative agreement is to be expected due to the finite range of spin
interactions $j(r)$ and the addition of the repulsive potential~$U(r)$, we
invert the canonical relations $\langle e \rangle(\beta, \bfalpha)$ and
$\langle \bfP \rangle(\beta, \bfalpha)$ from Eqs.~\eqref{eq:Avgp} and
\eqref{fc:e}. We thus obtain in Fig.~\ref{fig:constp} the magnetization
computed in the canonical ensemble as a function of ($E$, $\bfP)$ or more
conveniently ($\beta$,$\bfP$). For finite $\bfP$ continuous branches of solution
relate the disordered state at high temperature to the homogeneous magnetized
state at low temperature. These branches resemble very much those obtained
from the microcanonical simulations (Fig.~\ref{fig:compare_fc}). The
zero-momentum case is very peculiar: there is a finite range of energy, or
$kT$, where no solutions with homogeneous magnetization exist. Because of
finite-size effects, this peculiarity could not be captured in the
microcanonical simulations. Also, from the data in Fig.~\ref{fig:compare_fc}
it is not yet clear whether the magnetization at $kT>0$ survives the limit of
large systems, $N\to\infty$. The transition shifts more and more to the left,
but more systematic investigation would be necessary to know whether it
converges to a nonzero critical temperature. As a first step we looked at
finite size effects, focusing on the ground state. We ran the annealing
protocol on ten systems of size $N=2000$, for respectively $K=0.1$ and $0.3$.
By increasing the system size, the fully ordered state is now replaced by an
inhomogeneous state with zero total magnetization, in a way similar to the
effect of increasing $K$ at a given system size (Fig.~\ref{fig:gs}b--d). The
crossover size is larger for smaller~$K$. Given that for $K=0$ no long-range
order is expected, these observations suggest that long-range order may not
survive in the thermodynamic limit for all~$K$. Obtaining the precise scaling
in $K$, temperature and system size would require a deeper analysis, both
theoretically and numerically, which is beyond the scope of the present work.

\section{Discussion and conclusion}

In order to discuss the above observations, let us recall that the Hamiltonian
can be interpreted as the sum of the kinetic and potential energies,
in which the peculiarity of the present model is entirely encoded in the
unusual relation $\dot\bfr_i =\bfp_i-K\bfs_i$ (see Eq.~\eqref{eq:hamiltonian2}
in section~\ref{definition}). To obtain low energy states, the system should
(i) minimize its kinetic energy, thus decrease all $|\dot\bfr_i|$ and generate
small $|\bfV|$ states; (ii) align its spins, thus favoring states with large
$|\bfM|$. However this is incompatible with the constraint $\bfP = N \bfV + K
\bfM = \bfzero$. In particular, because the constraint imposes $N|\bfV| = K
|\bfM|$, the spin alignment potentially leads to large kinetic energy. This
dynamical frustration is all the more important that $K$ is large. This is why
decreasing $K$ is a way to favor ordered states at low energy. At larger $K$,
the system selects zero magnetization states, at the price of generating bend
and splay in the spin field. Also the larger the system, the smaller are these
distortions. This is why only small $K$ and small size system exhibit
homogeneously ordered phases.

In summary, we have proposed in this paper a conservative model of
particles in which velocities and spins are coupled. Starting from a Lagrangian
formulation and deriving from it the Hamiltonian one, we obtained using the
symmetries of the problem that the total (generalized) linear momentum is
conserved but is no longer proportional to the center-of-mass velocity. We
then studied the effect of this important change on spin statistics and on
collective motion. Our main findings are that (i) collective motion sets in
starting from an immobile system, provided that the spins are able to align
ferro-magnetically as observed in particular in the fully connected geometry,
or in small enough systems, and (ii) the parameter $\bfalpha =
\left<\bfV\right>$ thermodynamically conjugated to linear momentum acts as an
external magnetic field on the magnetization.

More generally, the main interest of the present model is to show that
collective motion, albeit perhaps of a nonstandard type, is possible even for
conservative models, provided that spins and velocities are coupled. Although it
is hard to imagine a physical realization of the present model, the latter
however has the virtue of showing that, at least at a conceptual level, energy
dissipation is not a necessary ingredient for collective motion. Also, in a
spirit similar to that of~\cite{Solon:2013vr} it brings the transition to
collective motion on a theoretical playground where a number of tools have been
developed to characterize phase transitions.

The present paper aimed at introducing the model, discuss its symmetries and
the crucial role of the broken Galilean invariance. As such, it remains very
preliminary. We have only looked at the mean field scenario, and illustrated
the model behavior on a few MD simulations. A number of perspectives can be
mentioned. Obviously one would like to investigate more systematically the
phase diagram of the system and the existence of a phase transition in
the infinite size limit. Also, it would be interesting to make progress in the
direction of the theoretical analysis of the model in finite dimension. In
particular two limits of interest are $K\rightarrow 0$, which corresponds to
the intricate physics of the XY-model and its Berezinskii-Kosterlitz-Thouless
transition, and $\eta \rightarrow 1$, where Galilean invariance is recovered.
Perturbative approaches using $K$ and/or $\eta$ as small parameters could
be a way to tackle this theoretical analysis. Besides, we have not yet explored
the possibility that the model might be invariant under a more complicated
transformation than the Galilean one, in analogy to the Lorentz transform that
arises in the context of electromagnetism. Whether such a transformation exists
is however far from clear; in electromagnetic systems, the invariance under the
Lorentz transform is only obtained when considering relativistic mechanics,
while our model remains within the realm of Newtonian mechanics.

\ack Interesting discussions with J.-L. Barrat on the
Mermin-Wagner theorem are acknowledged.

%

\appendix

\section{Computation of the mean-field partition function}

In this appendix, we provide the detailed derivation of the mean-field free
energy $\mathcal{F(\bfu)}$ given in Eq.~\eqref{eq:def-Fu}. Starting from
Eq.~\eqref{eq:Ztheta}, we need to compute the following integral,
\begin{equation}\label{eqA:Ztheta}
  Z_\theta = \int d\theta^{N} \, e^{\frac{1}{2} N\beta \bfm^2 + N\beta\bfm \cdot (K \bfalpha + \bfh)}
\end{equation}
To write the exponential function as a linear function of $\bfm$ we use the
Hubbard--Stratonovich transformation,
\begin{equation}
  \label{eqA:gauss}
  e^{b^2/2a} = \sqrt{\frac{a}{2\pi}}\int^\infty_{-\infty} du\, e^{-\frac{1}{2} a u^2 + bu}.
\end{equation}
Considering for instance the $x$-component, and setting $b=N\beta m_x$ and $a=N\beta$, one 
gets
\begin{align}
  e^{\frac{1}{2} N\beta m_x^2} = \sqrt{\frac{N\beta}{2\pi}}\int^\infty_{-\infty} du_1\, e^{- \frac{1}{2} N
\beta u_1^2 +N\beta m_x u_1}.
\end{align}
Combining the two directions $x$ and $y$ yields for the partition function,
using the notation $\bfu =(u_1,u_2)$,
\begin{equation} \label{eqA:Ztheta2}
  Z_\theta =\frac{N\beta}{2\pi} \int du_1 du_2 d\theta^N \,
    e^{-N\beta(\frac{\bfu^2}{2}- \bfgamma(\bfu) \cdot \bfm)},
\end{equation}
where
\begin{equation} \label{eqA:def-gamma}
  \bfgamma(\bfu)  \equiv  K \bfalpha +  \bfh + \bfu.
\end{equation}
Up to a shift on the variable $\theta_i$ (that is irrelevant since integration
is on the circle), the scalar product $\bfgamma(\bfu) \cdot \bfm$ can be
expressed as
\begin{equation}
  \bfgamma(\bfu) \cdot \bfm = \frac{1}{N} \sum_i \bfgamma(\bfu) \cdot \bfs_i
  = \frac{1}{N} \sum_i \gamma(\bfu) \cos\theta_i
\end{equation}
where $\gamma(\bfu) = |\bfgamma(\bfu)|$. We end up with
\begin{equation}
  Z_\theta = \frac{N\beta}{2\pi} \int du_1 du_2\,  e^{-N\beta \bfu^2/2}
    \prod_{i=1}^N \int_{-\pi}^{\pi} d\theta_i\, e^{\beta\gamma(\bfu)\cos\theta_i}
\end{equation}
Denoting $I_0(x)$ the modified Bessel function of order $0$:
\begin{equation}
  I_0(x) = \frac{1}{2\pi} \int_{-\pi}^{\pi} d\theta\, e^{x\cos\theta}
\end{equation}
we finally get Eqs.~\eqref{eq:Ztheta3} and \eqref{eq:def-Fu} of the main text.

\providecommand{\newblock}{}

\end{document}